\documentclass[aip,pof,reprint]{revtex4-1} 
\usepackage{graphicx} 
\usepackage[usenames]{color}
\usepackage{amsmath,amssymb}
\usepackage{gensymb}
\usepackage{mhchem}
\usepackage{balance} 
\usepackage{lastpage}
\usepackage{lipsum}
\usepackage[format=plain,justification=raggedright,singlelinecheck=false,font=small,labelfont=bf,labelsep=space]{caption} 
\usepackage[perpage]{footmisc}
\begin{document} 

\title{\noindent\LARGE{\textbf{Soft particles at a fluid interface}}
\vspace{0.6cm}}
\author{\noindent\large{\textbf{Hadi Mehrabian}}}
\affiliation{Physics of Fluids Group and J. M. Burgers Centre for Fluid Dynamics, University of Twente, P.O. Box 217, 7500 AE Enschede, The Netherlands}

\author{\noindent\large{\textbf{Jens Harting}}}
\affiliation{ Department of Applied Physics, Eindhoven University of Technology, P.O. Box 513, 5600 MB Eindhoven, The Netherlands}
\affiliation{Physics of Fluids Group and J. M. Burgers Centre for Fluid Dynamics, University of Twente, P.O. Box 217, 7500 AE Enschede, The Netherlands}

\author{\noindent\large{\textbf{Jacco H. Snoeijer $^{\ast}$}}} \vspace{6pt} 
\affiliation{Physics of Fluids Group and J. M. Burgers Centre for Fluid Dynamics, University of Twente, P.O. Box 217, 7500 AE Enschede, The Netherlands}
\affiliation{ Department of Applied Physics, Eindhoven University of Technology, P.O. Box 513, 5600 MB Eindhoven, The Netherlands}

\date{\today} 

\begin{abstract}
Particles added to a fluid interface can be used as a surface stabilizer in the
food, oil and cosmetic industries. As an alternative to rigid particles, it is
promising to consider highly deformable particles that can adapt their
conformation at the interface. In this study we compute the shapes of soft
elastic particles using molecular dynamics simulations of a cross-linked
polymer gel, complemented by continuum calculations based on linear elasticity.
It is shown that the particle shape is not only affected by the Young's modulus
of the particle, but also strongly depends on whether the gel is partially or
completely wetting the fluid interface. We find that the molecular simulations
for the partially wetting case are very accurately described by the continuum
theory. By contrast, when the gel is completely wetting the fluid interface the
linear theory breaks down and we reveal that molecular details have a strong
influence on the equilibrium shape. 
\end{abstract} 

\maketitle

\section{Introduction}
An important application of particle stabilized fluid interfaces goes back to the
importance of retaining the dispersivity of emulsions \citep{bib:binks-horozov:2006,bib:sagis:2011}. The dispersion process
results in a large interfacial area and hence high interfacial energy. Being
energetically unfavourable, dispersed phases will eventually coarsen and form
phase separated volumes of fluids \citep{bib:taylor:1998}.
A traditional way to stabilise the dispersions with respect to phase separation
is to use surfactants. As an alternative for surfactants, an interface can also
be made kinetically stable by adding solid particles \citep{bib:binks:2002,bib:binks-horozov:2006}. Particle-stabilised
emulsions, also called Pickering emulsions \cite{bib:ramsden:1903,bib:pickering:1907} are metastable since the
particles anchor to the interfaces much more strongly than the surfactant
molecules. For example, particles as small as a few tens of
nanometres have a desorption energy as high as $10^3 $ to $ 10^4k_BT$, while it
is around $10k_BT$ for surfactant molecules \citep{bib:binks-horozov:2006,bib:davies-krueger-coveney-harting:2014}. 

As an alternative to solid particles, soft particles have recently attained
attention as stabilizers for emulsions \citep{DeshmukhSoftHard}. The shape of the particles can adapt to
the interface and depends on the interplay of the molecular interactions
between particle and fluids as well as the elastic properties of the particle
itself. The most important types of relevant particles are cross-linked polymer
networks (microgels) \citep{Zhang99,Brugger2009,Saunders1999,Saunders2009} and
ligand or polymer grafted nanoparticles \citep{Nelson2015,Tay2006a,Isa2011}.
Combinations of both types are so-called core-shell
microgels \citep{Karg2006,Khan2007}. The desorption energy of micron-sized
microgel particles at an interface was found to increase even up to $10^6
k_BT$ \citep{Monteillet2014}, making them very promising as
stabilizers \citep{Schmidt2011,Li2013a,Destribats2013,Destribats2011a,Destribats2014}.

The presence of a network of interconnected polymer chains makes these
particles deformable under an external forcing. They can show the properties of
both the individual chains as well as the particles with well-defined
boundaries~\citep{Lyon2012}. This highlights the role of the interfacial
tension of soft particles in their deformation at a fluid interface.
Understanding the details of the adsorption of soft particles to fluid
interfaces is highly complex, as it involves e.g. the cross-linking of the
polymer network, temperature and pH value~\citep{Destribats2011a}. However,
obtaining a fundamental understanding of the conformation of these objects at a
fluid interface in terms of the macroscopic and microscopic parameters is
crucial for exploiting them in many practical applications.

\begin{figure*}[h!t]
\centering
\includegraphics[height=9cm]{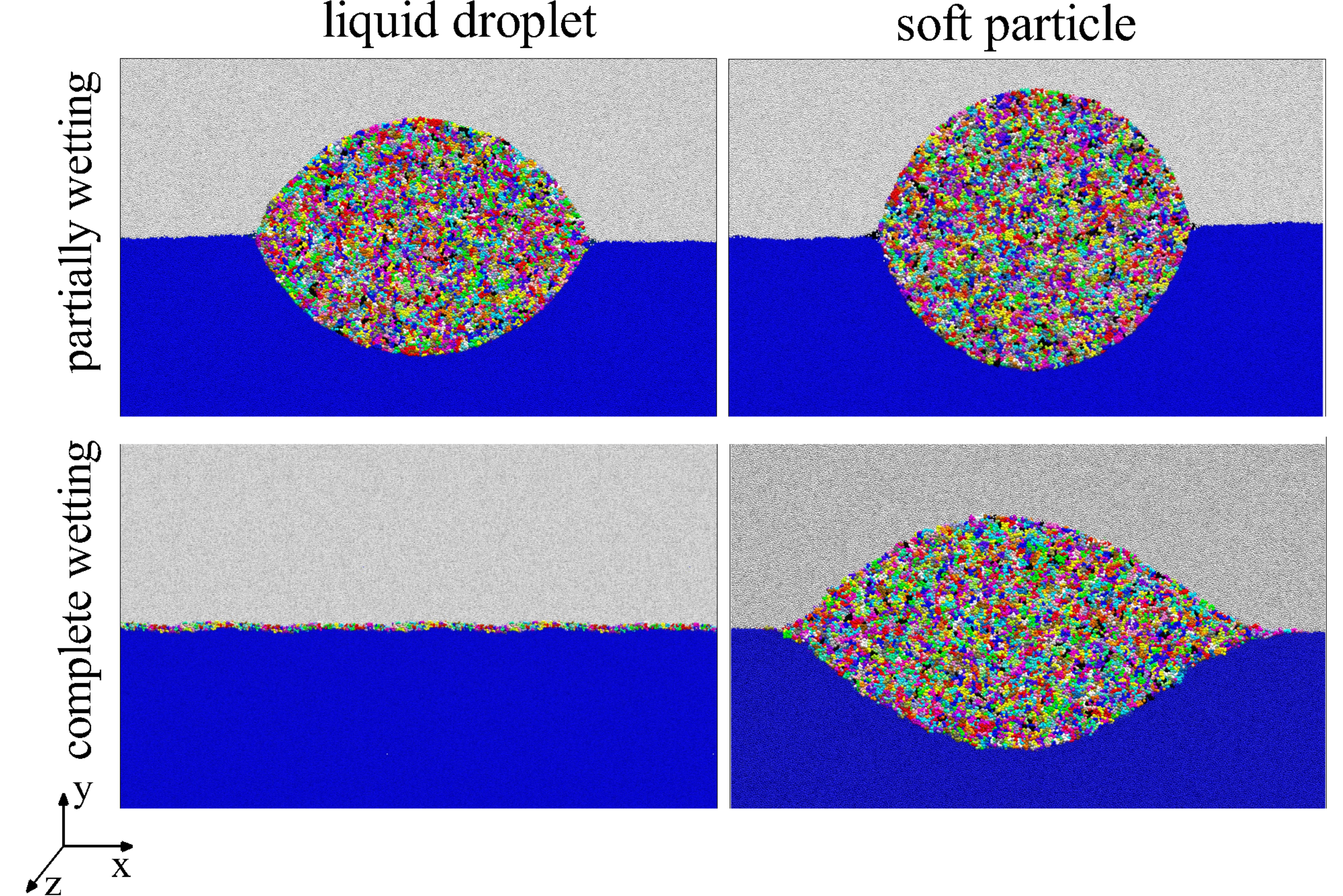}
\caption{Snapshots from molecular dynamics simulations of soft particles at a fluid-fluid interface. The key parameters of this study are the elastic modulus and the wettability of the particles. The left panels show the outcome for a liquid drop, consisting of a non-cross-linked polymer, respectively for partially wetting (top) and complete wetting (bottom). The right panels show the corresponding particle shapes when the polymers are cross-linked and have a finite elastic modulus. The reason for small asymmetry between left and right hand side of the soft particle in the complete wetting regime is explained in section~\ref{s:wetting}.} 
\label{f:collage}
\end{figure*}

Most experimental studies on the interaction of soft particles and fluid
interfaces focus on the behaviour of a single microgel particle at a fluid
interface \citep{Richtering2012,Geisel2012} or bulk properties of microgel
covered interfaces
\citep{Garbin2015,Geisel2014,Monteux2010,Pinaud2014d,Deshmukh2014a} whose key observation is that the microgel particles take a core-corona (also called fried-egg) shape at a fluid interface. There are some challenges when studying the deformation of a particle at a fluid interface: due to the small size of microgel particles and their reflective index being very close to that of the surrounding solvent, imaging is difficult \cite{Monteillet2014}. In addition, soft particles are very sensitive to
external stimuli such as pH or temperature change.  There are few numerical simulations of polymer grafted nanoparticles using molecular dynamics. For
example, Schwenke et al. \citep{Schwenke2014} have studied the particle conformation and particle-particle interaction of polymer coated nanoparticle in different solvents. Udayana et al.
\citep{Udayana2010} have studied the adsorption/desorption energy of polymer coated nanoparticles to/from an interface.  Lane et al.  \citep{Lane2010} have
investigated the distribution of polymer chains. None of these works has discussed the shape of the soft particles in terms of their macroscopic
properties.

From a theoretical point of view, it is appealing to investigate the shape of
soft particles using a continuum framework based on linear elasticity. As the
particle becomes increasingly soft, however, such a theory must also account
for interfacial forces. The interplay of capillarity and elasticity was
recently investigated in great detail in the context of adhesion of particles
\cite{Cao2014,Carrillo2010,StyleNatComm2013,SalezSM2013} or the wetting of liquid droplets on highly deformable substrates
\cite{Lubbers2014a,Limat12,Style2012,SBCWWD13,BostwickSM14,dervaux}. In these studies the governing parameter was
found to be the elasto-capillary length $\gamma_s/E$, comparing the surface tension
of the solid $\gamma_s$ to its Young's modulus $E$. If the particle (or
droplet) size $R$ is large with respect to $\gamma_s/E$, it can be considered
effectively rigid -- except in a small region near the contact line. In the
opposite limit where the particle is small with respect to the elasto-capillary
length, the elasticity is so weak that the elastic medium can effectively be
considered as a liquid with a surface tension. Extending this point of
view to the adsorption of soft particles at interfaces, one thus expects two
limiting cases~\cite{Monteillet2014,Style2015Adsorption}: one is the behaviour of a
perfectly rigid particle at an interface (governed by the Young contact angle),
while the soft extreme of vanishing elasticity corresponds to a liquid droplet
at a fluid-fluid interface (governed by the Neumann contact angles \cite{bookdeGennes}).
It has remained unclear to what extent linear elasticity can describe the shape
of soft particles at fluid-fluid interfaces, in particular when the surface
tension of the particle is relatively low~\cite{Style2015Adsorption}.

In this paper we quantify the equilibrium shapes of soft particles at
fluid-fluid interfaces, by combining molecular dynamics simulations and exact
solutions derived from linear elasticity. Previously, the molecular dynamics
method has been used to study the interaction between a liquid droplet or a
solid nanoparticle with a fluid interface  \citep{Bresme2000,Razavi2013a}, and
the interaction between soft materials made of inter-connected polymer chains,
with liquid or solid surfaces
\citep{Leonforte2011a,Carrillo2010,Cao2015,Cao2014}.  Similarly, our particles
consist of a cross-linked polymeric liquid and are adsorbed at a fluid-fluid
interface. By varying the molecular interactions and cross-linking density, we
can explore a broad range of Young's moduli and interfacial energies.
Snapshots of typical simulations are shown in Fig.~\ref{f:collage}, where the
individual polymer chains can be identified by their color. Apart from an
expected dependence on the Young's modulus, governed by the dimensionless
parameter $\gamma_s/(ER)$, our key finding is that one needs to distinguish the
cases where the polymer is partially or completely wetting. The left column of
Fig.~\ref{f:collage} shows simulations of the polymer \emph{liquid} (i.e.
without cross-linking), respectively for partial wetting (upper panel) and
complete wetting (lower panel). The right column shows equivalent systems, but
now for \emph{soft solid particles} that are cross-linked and thus exhibit a
finite elastic modulus. We find that the partially wetting systems are very
accurately described by linear elasticity, as long as the solid surface tension
$\gamma_s$ is comparable or larger than that of the fluid-fluid interface. In
the case of complete wetting, by contrast, the interfacial forces favor a state
where the gel covers the entire interface but this is prohibited by the network
elasticity. This leads to an intricate elasto-capillary balance, where the
elastic response is highly nonlinear. In  addition, we observe an important
influence of molecular details of the cross-linking, signalling a breakdown of
continuum theory. 


%
The paper is organised as follows: First, we discuss the analytical continuum
theory, which is limited to the regime of small deformations, in section
\ref{s:theory}. Details of the molecular dynamics simulations are presented in
section~\ref{s:md} and our simulation results are presented in section
\ref{s:results}. Here we provide a detailed comparison to the continuum
predictions and explore the regime of complete wetting. The paper closes with a
discussion where we also comment on the implications of our findings for
microgel particles (section~\ref{s:discussion}).

\begin{figure}[h]
\centering 
  \includegraphics[height=7.0cm]{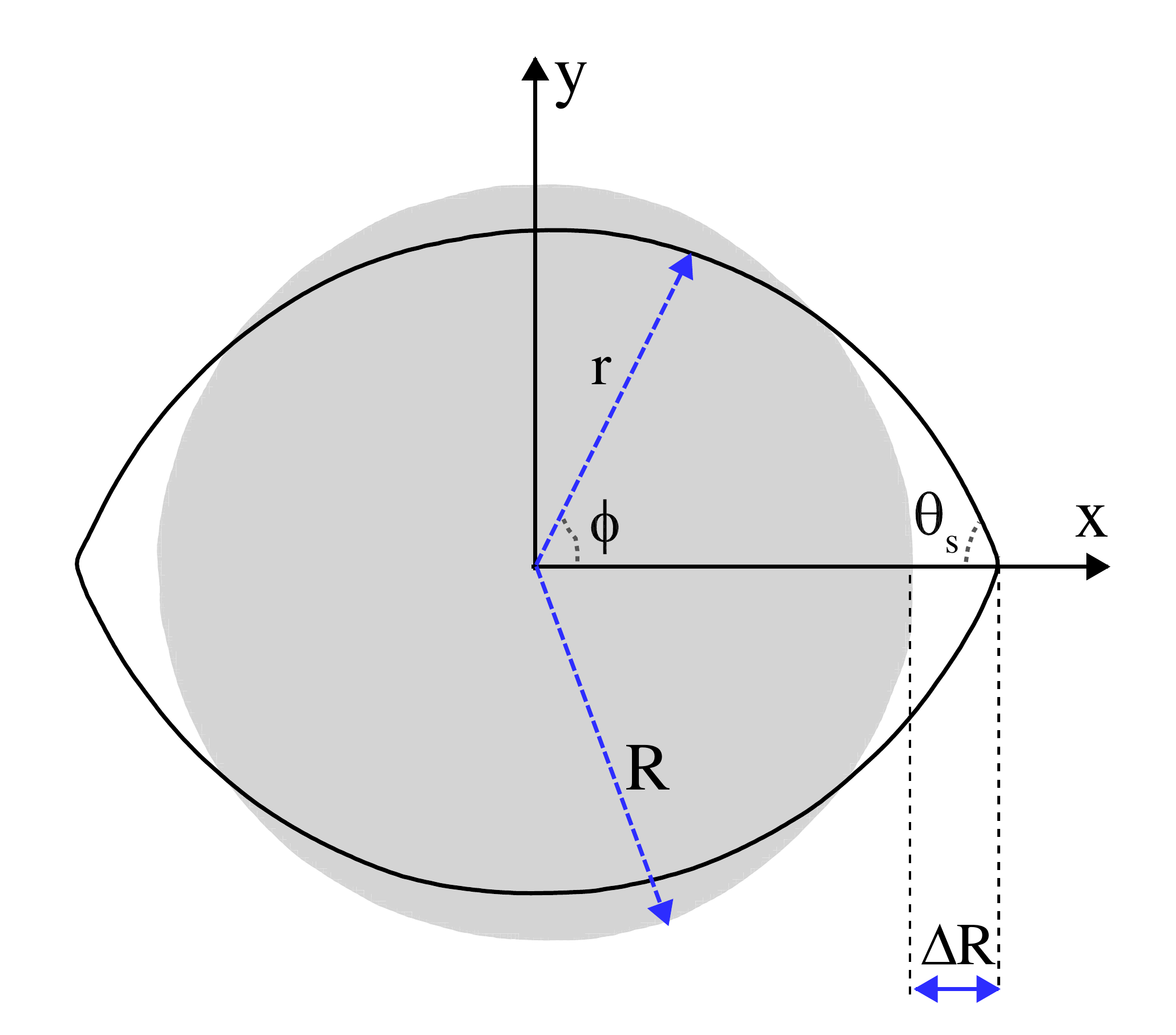}
  \caption{Schematic shape of the undeformed (grey circle) and
deformed particle (solid line). $(r,\phi)$ denotes the components of the polar
coordinates. The fluid-fluid interface is located along the x-axis, at $\phi=0$
and $\pi$. $R, \Delta R$ and $\theta_s$ denote the particle's radius, its
maximum deformation and contact angle.}
\label{f:schem}
\end{figure}




\section{Macroscopic thermodynamic formulation}\label{s:theory}

In this section we pose the problem of an elastic particle at an interface from a macroscopic thermodynamic perspective. The relevant dimensionless parameters are identified and the shape of the deformed particle is computed in the framework of linear elasticity. This provides a benchmark for the case of weakly deformed particles, and outlines the key issues that will later be explored by molecular dynamics for large particle deformations. This study is done in a two-dimensional setup. It is expected that the two-dimensional simulations capture the salient features of the three-dimensional configuration because the main physics of the problem happens at the vicinity of the contact lines except that the global deformation in three-dimensions will be slightly less than in the two-dimensional case due stress in the azimuthal direction.

\subsection{Dimensionless parameters}

We consider a two-dimensional elastic particle (Young's modulus $E$) that is placed at a liquid-liquid interface (surface tension $\gamma_b$), as sketched in Fig.\ref{f:collage}. For simplicity, the study is limited to the cases where the solid particle has an equal surface tension with the two liquid phases, simply denoted $\gamma_s$, so that the particle will develop a symmetric shape \footnote{In case the surface energy varies with the amount of stretching, one needs to distinguish between surface free energy and surface stress for the liquid-gel interface due to the Shuttleworth effect \cite{Shuttleworth50}. Since here we consider a system with two identical liquids, this effect does not induce a forcing tangential to the particle surface \cite{weijs2013} and our calculation is valid --  in the remainder we simply refer to surface tension, which for the liquid-gel interface must be seen as the surface stress.}. 
The reference state of the particle is a circle of radius $R$, which is for example achieved by cross linking a liquid drop that is fully immersed in one of the two liquids. We assume the particle to be incompressible (Poisson ratio $\nu=1/2$), although the results are easily generalised to an arbitrary Poisson ratio. In analogy to the case of liquid drops on deformable substrates \cite{Limat12,Style2012,Lubbers2013}, there are two dimensionless parameters that characterise the problem, namely 
\begin{equation}
S = \frac{\gamma_s}{ER}, \quad {\rm and} \quad \gamma = \frac{\gamma_s}{\gamma_b}.
\end{equation}
The first term is called the ``softness" of the particle and compares the elasto-capillary length $\gamma_s/E$ (based on the solid surface tension), to the particle radius. The second term is the ratio of surface tensions, controlling the wetting conditions.
Before turning to a detailed analysis, it is instructive to discuss the extreme limits of the softness $S$. For $S=0$, the particle can be considered as undeformable (large $E$ or $R$) and one recovers the usual case of a rigid spherical colloid at an interface. The limit $S=\infty$ corresponds to a particle without any rigidity, as is achieved for a polymeric drop without any cross-linking. Hence, in this limit one expects to recover the ``liquid lens" floating at the interface \cite{bookdeGennes}. The shape of such a lens is governed by the ratio of surface tensions $\gamma$, and its contact angle $\theta_s$ follows from the Neumann triangle (Fig.\ref{f:schem}). Given the symmetry of our problem, where the gel has an equal surface tension with the two liquids, the Neumann balance of surface tensions reads

\begin{eqnarray}\label{eq:neumann}
\gamma_b = 2\gamma_s \cos \theta_s \quad \Longrightarrow \quad \theta_s=\arccos\left(\frac{1}{2 \gamma}\right). \label{qs}
\end{eqnarray}
Quantifying the maximum drop deformation $\Delta R$ as defined in Fig.\ref{f:schem}, we find that 

\begin{eqnarray}\label{eq:dRNeumann}
\frac{\Delta R_{drop}}{R}=\sqrt{\frac{\pi}{2\theta_s- \sin 2\theta_s}} \sin \theta_s -1 \quad {\rm for} \quad S \rightarrow \infty. \label{lens_dR}
\end{eqnarray}
Note that such a drop exhibits a transition from \emph{partial} wetting to \emph{complete} wetting when the surface tension ratio $\gamma \rightarrow 1/2$: at smaller values of $\gamma$ the Neumann balance (\ref{eq:neumann}) cannot be satisfied. The transition corresponds to $\theta_s \rightarrow 0$ for which Eq.~(\ref{eq:dRNeumann}) gives a divergence of $\Delta R/R$, signalling the onset of a wetting layer.


%
%

\subsection{Small deformations: linear elasticity}
In the limit of small deformations, one can resolve the particle shape using linear elasticity theory. The two-dimensional problem consists of a disk of radius $R$ that is deformed under the influence of an external stress at the free surface. This stress, or traction, has two contributions. First, the liquid-liquid surface tension pulls on the particle at $\phi=0$ and $\pi$, where we use polar coordinates as defined in Fig.~\ref{f:schem}. This traction is described by a perfectly localised force per unit length, using a Dirac $\delta$ distribution. Second, the deformation changes the curvature of the particle-liquid interface, and as such induces a capillary pressure that acts as an additional traction $\gamma_s \kappa$. In this expression $\gamma_s$ is the surface stress (here simply denoted as surface tension), while $\kappa$ is the extra curvature of the solid-liquid interface due to the deformation, which to linear order in the radial displacements $u_r$ reads

\begin{equation}
\kappa = \frac{1}{R^2}\left( u_r  + \frac{\partial^2 u_r}{\partial \phi^2} \right),
\end{equation}
to be evaluated at the disk boundary $r=R$. The importance of this solid Laplace pressure has been highlighted in several recent papers involving very soft gels \cite{MPFPP10,Jerison11,MDSA12,ParetkarSM14}, and will also be apparent in the present work. The total traction boundary condition thus becomes

\begin{eqnarray} \label{eq:bc1}
\sigma_{rr}(r=R,\phi) &=& \frac{\gamma_b}{R} \left[ \delta(\phi) + \delta(\phi-\pi)  \right] +  \gamma_s \kappa, \\ 
\sigma_{r\phi}(r=R,\phi) &=& 0, \label{eq:bc2}
\end{eqnarray}
where the latter expresses the no-shear stress boundary condition. Interestingly, the traction depends on the displacement. As a consequence, the resulting tractions and displacements have to be determined self-consistently. 

%

\begin{figure}[!h]
\begin{center}
\includegraphics[width=7.5cm]{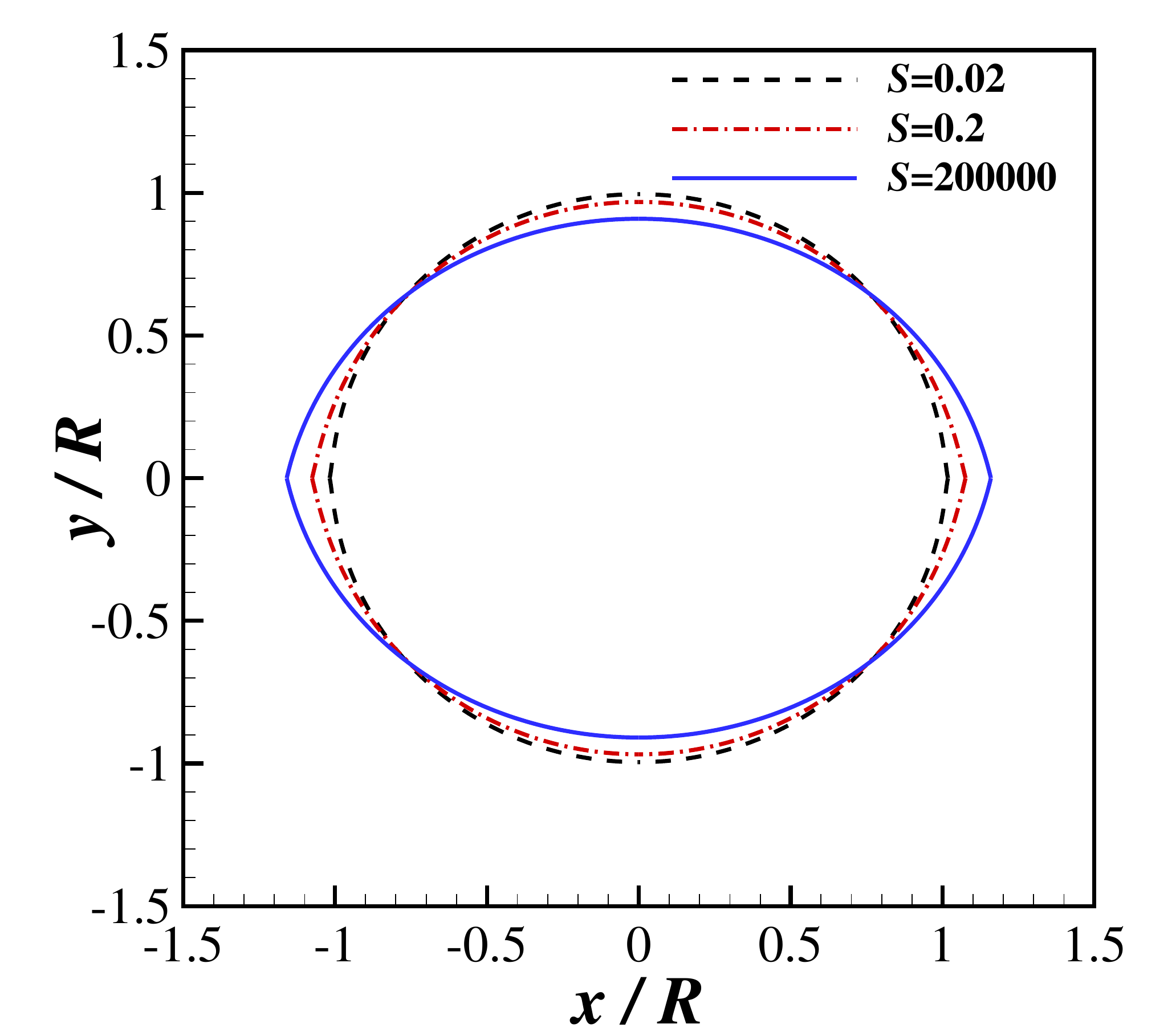}
\caption{ Deformation of a 2D particle at an interface, predicted by equation (\ref{eq:disk}) for $\gamma=2.0$ and $S=200000, 0.2$ and $0.02$. The contact angle $\theta_s=75.5^\circ$ for all cases, dictated by the Neumann balance. 
}
\label{f:theory}
\end{center}
\end{figure}

This two-dimensional problem can be solved using the Airy stress function formalism \cite{Musk}, and the complete derivation is given in the Appendix. The result is obtained using a series expansion in polar coordinates, based on the classical solutions by Michell. The resulting shape of the deformed particle is given by (cf. Appendix)

\begin{eqnarray}\label{eq:disk}
\frac{r_{\rm p}(\phi)}{R} = 1 + \frac{3S}{2\gamma \pi} \sum_{k=1}^\infty    \frac{4k \cos 2k\phi}{(4k^2-1)(1+3 S k)}, 
\end{eqnarray}
which indeed depends on both the dimensionless softness $S$ and the ratio of surface tensions $\gamma$. A similar expression was derived for axisymmetric particles in \cite{Style2015Adsorption}. Some examples of the deformed particles are shown in Fig.~\ref{f:theory}. Importantly, the particle exhibits a finite contact angle $\theta_s$, for \emph{all} values of the softness $S$, as can be seen from the kinks at the position where the fluid-fluid interface applies its force. In the theory, the angle $\theta$ is even independent of the elastic properties of the particle. This can be inferred from the large $k$-asymptotics of (\ref{eq:disk}), which gives a slope discontinuity in $dr_p /\partial \phi$ of magnitude $1/\gamma$. In terms of the contact angle this gives 

\begin{equation}\label{eq:neumannsmall}
\theta_s = \frac{\pi}{2} - \frac{1}{2\gamma},
\end{equation}
which is the small slope, or large $\gamma$, equivalent of (\ref{eq:neumann}). Hence, the continuum theory predicts that the contact angle is determined solely by the surface tensions, satisfying Neumann's law, regardless of the stiffness. This is in perfect analogy to results previously obtained for liquid drops on soft solids. The Neumann’s angle also appear for axisymmetric soft particles \cite{Style2015Adsorption}.

The maximum radius of the particle is attained at $\phi=0$ and $\pi$, for which the series (\ref{eq:disk}) can be summed to the explicit form

\begin{equation}\label{eq:DRS}
\frac{\Delta R}{R} = \frac{3S\left[ 2-2\Psi(1/3S) -4 \ln(2) - 2\gamma_E - 3S \right]}{\gamma(9S^2-4)\pi}.
\end{equation}
Here $\gamma_E$ is the Euler-Mascheroni constant and $\Psi$ is the digamma function. As expected, the deformation $\Delta R/R$ vanishes for a rigid particle ($S=0$) and saturates to a finite value in the very soft limit ($S\rightarrow \infty$). In the soft limit, we find $\Delta R/R \rightarrow 1/(\pi \gamma)$, which agrees with (\ref{eq:dRNeumann}) for large $\gamma$. It the rigid limit, we obtain [using $\Psi(x) \simeq \ln x $]:

\begin{equation}\label{eq:log}
\frac{\Delta R}{R} \simeq \frac{3}{2 \pi } \, \frac{S}{\gamma} \ln\left(\frac{a}{S} \right) = \frac{3}{2 \pi }\,  \frac{\gamma_b}{ER} \ln\left(\frac{aER}{\gamma_s} \right), \: {\rm for}  \quad  S\ll 1.
\end{equation}
where the numerical constant $a=\frac{4}{3}e^{\gamma_E-1}$. The deflection thus increases linearly with softness, involving the fluid-fluid surface tension $\gamma_b$, with logarithmic corrections that involve the the surface tension of the solid $\gamma_s$. Similar scaling laws for contact line deflections were obtained for drops on gels \cite{Limat12,Lubbers2013,dervaux}. Hence, the surface tension of the solid $\gamma_s$ is critical to achieve a finite elastic deformation in the macroscopic theory: otherwise, the deformation would be (logarithmically) divergent. In addition, $\gamma_s$ determines the boundary condition for the contact angle, following Neumann's law. We will comment later how this picture must breakdown in the regime of complete wetting, for which the surface tensions cannot achieve the Neumann balance.

The results from the molecular dynamics simulations will be directly compared to the shape (\ref{eq:disk}), the Neumann contact angle prediction (\ref{eq:neumann}), as well as the maximum extension $\Delta R$. For the latter, it is convenient to normalise the extension (\ref{eq:DRS}) by its limiting value achieved for a liquid droplet at large $\gamma$ i.e. $1/(\pi\gamma)$, so the result only depends on the softness $S$ (not on $\gamma$):

\begin{equation}\label{eq:DRS2}
\frac{\Delta R}{\Delta R_{drop}} = \frac{3S\left[ 2-2\Psi(1/3S) -4 \ln(2) - 2\gamma_E - 3S \right]}{(9S^2-4)}.
\end{equation}

\subsection{Observations and questions}
The macroscopic theory provides quantitative predictions, for the shape and the contact angle, that will be tested using molecular dynamics simulations. However, the simulations are even of more interest when the linear theory breaks down, and we therefore point out some important limitations of linear elasticity. First, the analysis only applies for large $\gamma$, for which according to (\ref{eq:neumannsmall}) the Neumann angle $\theta_s$ is always close to $\pi/2$. The reason for this is that in the linear elasticity calculation we have evaluated the curvature $\kappa$ to lowest order in $\partial u_r/\partial \phi$ -- keeping nonlinear terms prevents the solution by Fourier expansion. As a consequence, the theory cannot capture the wetting transition at $\gamma =1/2$, for which the angles are no longer small. Of particular interest is the force balance in the vicinity of the contact line. Contrarily to the linear theory, for the situation of complete wetting the Neumann balance cannot be satisfied and an elastic contribution must emerge to balance the localised contact line force. How does the wetting transition appear in the case of finite elasticity?

\section{Molecular dynamics simulation}\label{s:md}

We study the particles in a quasi two-dimensional setup, allowing to simulate a
large particle at a less computational cost as well as a direct comparison to
the elasticity theory described above. The setup is prepared in three main
steps. First a fluid-fluid interface is made.  Then a cylindrical particle is
built from polymeric chains and finally the particle is inserted at the
interface and the system is equilibrated.  In addition, two auxiliary setups
are needed to measure the Young's modulus of the particle and the
liquid-particle surface tensions. Below, these steps will be outlined in
detail.
We use the molecular dynamics method implemented in the GROMACS 5.0.2 \citep{Pronk2013,Hess2008,VanDerSpoel2005a,Berendsen1995} software to do the numerical simulations. Integration of Newton's equations is done using the leap-frog algorithm. Neighbour searching is done using verlet dynamic lists.  Visualization of the results is done using the VMD package \citep{Humphrey1996}.

\subsection{Interface modelling}
To keep the model as simple as possible, we use the modified shifted and truncated Lennard-Jones potential \citep{Stecki1995} to model the interaction between each pair of particles,
\begin{equation}\label{LJ}
U = \left\{
  \begin{array}{ll}
    4\epsilon \left[ \left(\frac{d}{r}\right)^{12}- \left(\frac{d}{r_c}\right)^{12} +\alpha\left[\left(\frac{d}{r_c}\right)^6 - \left(\frac{d}{r}\right)^6\right] \right]& \quad \mbox{if $r \leq r_c$}\\
    0 & \quad \mbox{if $r > r_c$.}\\
  \end{array} 
\right.
\end{equation}
Here $r$ is the distance between two particles, $d$ is the repulsive core
diameter fixed to $0.34$ nm, $\epsilon$ is the depth of the potential
minimum set to 3 kJ/mol, and $r_c$ is the the cut-off radius
equal to $5d$.  The characteristic time of the system is $\tau=\sqrt{\frac{m
d^2}{\epsilon}}=0.6 ps$ and $m=10amu$ is the particle mass. Using this set of
parameters, the surface tension of the liquid-liquid interface would fall in
the range between 0 and 73 mN/m, which are typical experimental values. In the
remainder of the paper, we will not use SI units, but reside to either
plots nondimensionalized by the particle radius, or use $\tau$, $d$, and $\epsilon$ as the reference values
for the time, length, and energy, respectively.
The coefficient $\alpha$ is set to 1 for two identical Lennard-Jones particles
and it is smaller than 1 when two particles are of different species. This
naturally leads to phase separation, where $\alpha$ controls the degree of
miscibility and hence the surface tension. 
In this study there are three phases and hence three values for $\alpha$ are
required. Since only symmetric liquid-particle interactions are considered,
both fluids have the same interaction parameter with the gel, which we call
$\alpha_p$. Therefore, only two values of $\alpha$ and $\alpha_p$ will be
varied to produce different values for the ratio of surface tensions $\gamma$.

The liquid-liquid interface is produced under the $NP_nT$ ensemble where $P_n$ denotes the pressure normal to the interface \citep{Zhang1995}. 
For keeping the temperature and normal pressure constant, Nose-Hoover temperature coupling\citep{Hoover1985} and Parrinello-Rahman pressure coupling \citep{Parrinello1981} methods are used, respectively.
Periodic boundary conditions are applied in all three directions. In our simulations, the produced liquid-liquid interface has approximately two million Lennard-Jones particles and its dimensions in $x$, $y$, and $z$ directions, as defined in Fig.~\ref{f:collage}, are $600d$, $300d$, and $14.7d$, respectively.

\subsection{Particle modelling}
The particle is made from interconnected polymeric chains. Each chain is
constructed using a coarse-grained model of beads and springs and consists of
32 monomers where two neighbouring beads interact via the following potential:

\begin{equation} \label{eq:FENE}
  U(r)=-\frac{1}{2} k_s R_{max}^2 \ln\left(1-\frac{r^2}{R_{max}^2}\right)+4\epsilon \left(\frac{d}{r}\right)^{12}.
\end{equation}
The first term on the right hand side is the so-called FENE (finite extensible nonlinear elastic) potential which controls the amount of the extension. The second term is the Lennard-Jones repulsion term that accounts for the reduced volume effect and prevents the collapse of the beads onto each other. We fix $ k_s=25 (\frac{\epsilon}{d^2})$ and $R_{max}= 1.5 d$. These values are chosen such that the simulation timesteps can be taken on the order of the pure Lennard-Jones fluid while making the link breakage energetically impossible \cite{kremer1990}.
By randomly cross-linking the polymeric chain, one creates a network of entangled polymers that exhibits elasticity in the long time limit. By increasing the density of the cross-linking, the rigidity of the network increases leading to a larger Young's modulus. %
All non-neighbouring beads of the chains interact with each other and also with the beads of the other chains through the Lennard-Jones potentials as described in the previous section. It is assumed that the particle is an isotropic material and the density of the cross-links is distributed uniformly over the particle volume. In addition, the number of cross-links per chain is kept fixed to distribute the cross-linkings over all chains. The interaction between the polymeric beads and the liquid particle follows the modified Lennard-Jones potential with the interaction parameter called $\alpha_p$. 

In order to make a two dimensional setup, we fix the size of the simulation box in z direction. The minimum depth of the simulation box is determined by two considerations. Lennard-Jones particles should not interact with themselves through the periodic boundaries. This condition is met by choosing the depth to be twice the cutoff radius. Since the cutoff radius is $5d$, $10d$ is sufficient to remove the Lennard-Jones self-interaction of particles. 
Second, polymeric chains should not interact with themselves through periodic boundaries. In our simulations, polymer chains consist of $N=32$ monomers which makes their average radius of gyration equal to $R_g= \sqrt{\frac{N d^2}{6}}=2.3d$. Therefore $2(R_g+ r_c) = 14.6d$ removes most of the self-interaction of the polymer chains.
This condition is only important for polymeric liquids which have a small number of cross-links since the forces due to FENE cross-links are much stronger than the Lennard-Jones interactions. Thus the dynamics of the gel is mainly determined by the FENE-links. In this study, the depth of the setup is chosen to be $14.7d$. In addition, there is a maximum value for the depth to avoid the capillary instability. For a radius of $R \approx 75d$, capillary instability happens when depth of the simulation box is approximately $475d$ which is far above the considered range for the depth. 
%

In order to create the polymeric particle, first a big chunk of polymeric chains is made. Then those chains are positioned inside a bath of solvent, which naturally leads to the formation of a (cylindrical) polymeric droplet. 
Based on the desired cross-linking density, the polymeric droplet is cross-linked using (\ref{eq:FENE}) and then relaxed to reach its final equilibrium size. In our simulations, the resulting particle has 250,000 beads and its radius is $75.3d \pm 0.3d$.

\subsection{Modelling of interface and particle}
After preparation of the equilibrated interface and cross-linked particle, the particle is inserted at the interface by carving out some volume of the liquid. 
It has been observed that a gap could be produced where the three phases are in contact which can be removed by increasing the pressure normal to the interface \cite{Bresme2000}. 
After the insertion, the system is equilibrated under the NVT ensemble for $20,000 \tau$. The Nose-Hoover thermostat is used to keep the temperature constant at 300 K. We use the time step $ dt=0.01 \tau $. After the equilibration, the shape of the particle is measured every $10,000\tau$ measurement steps. The criterion for convergence is that the shape of the particle in the three consecutive measurement steps is identical.


\begin{figure}[h]
\centering
  \includegraphics[height=7cm]{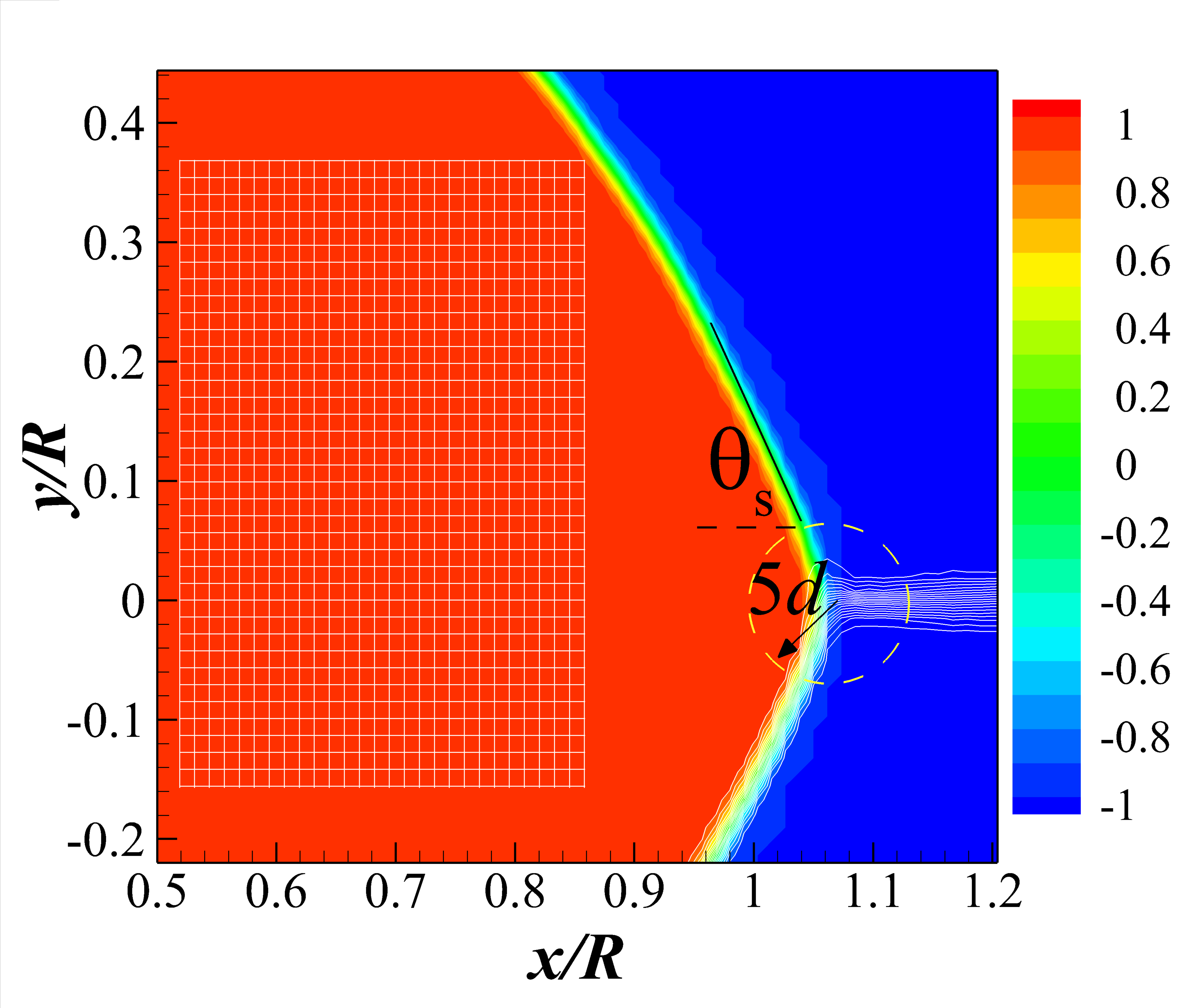}
\caption{Density contours $\psi$, as defined in  (\ref{eq:rho}). To obtain
$\psi$ contours, the density is computed inside the square boxes of size
$0.9d^2$ in the $x$-$y$ plane as depicted by the white grid. In addition, for
the measurement of the macroscopic contact angle $\theta_s$, a region of size
$5d$ is excluded from the contact point to exclude the effect of finite width
of the fluid interface (shown with the line contours).}
\label{f:density}
\end{figure}

The main goal of the study is to quantify the shape of the particles as a function of the various parameters. To obtain the shape from a molecular dynamics simulation, first the simulation data is extracted every $20 \tau$ and the simulation box is shifted in a way that the center of mass of the particle remains fixed in all output frames. Then the simulation box is divided into smaller square boxes with an area of $(0.9d)^2$ in the $x$-$y$ plane, as it is depicted by the white grid in Fig.~\ref{f:density}. Inside each box the densities of polymeric chains and two liquid phases are calculated for each frame. Next, the densities are averaged over 500 frames. Finally, iso density contours for each measurement step are obtained using \citep{Lacasse1998}
\begin{equation} \label{eq:rho}
\psi=\frac{\rho_s-(\rho_1+\rho_2)}{\rho_1+\rho_a+\rho_2},
\end{equation}
where $\rho_a$ and $\rho_b$ denote the density of two liquid phases $a$ and $b$ and $\rho_s$ denotes the density of the particle. A typical density contour is shown in Fig.\ref{f:density}.
Particles have four symmetric quadrants which can be used to increase the
amount of data available for the averaging process by a factor of four. To
obtain a macroscopic contact angle from the simulations, a small region on the
order of the interface thickness, here chosen $5d$, needs to be excluded. A
typical contact angle measurement is shown in Fig.~\ref{f:density}.

\subsection{Calibration} \label{s:calib}
In order to compare the results of the simulations with any macroscopic model,
it is required to calculate the Young's modulus $E$ of the particle and the
surface tensions $\gamma_s$ and $\gamma_b$. This calibration is outlined
below.

\subsubsection{Elastic modulus measurement.~~}
\begin{figure}[h]
\centering
  \includegraphics[height=7.5cm]{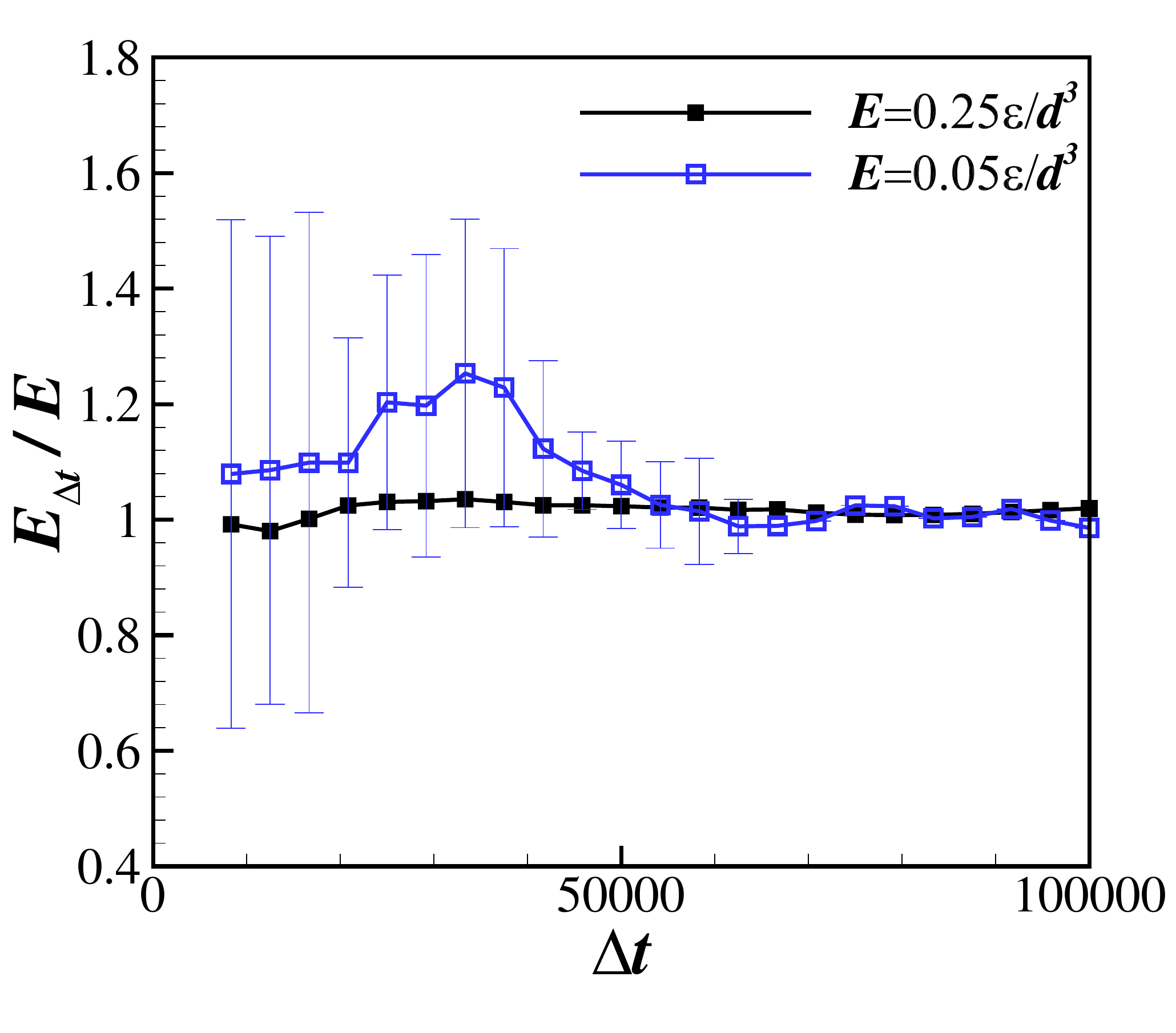}
\caption{Variation of the instantaneous Young's modulus $E_{\Delta t}$
(normalised by its time-averaged value $E$) with the interval of averaging, for
two systems of different $E$. Error bars show the order of the fluctuations for
each interval of the averaging, and give an indication of the convergence. For
smaller average elasticity, a longer period of averaging is required to measure
the Young's modulus.} 
\label{f:E}
\end{figure}

To measure the Young's modulus we prepare a cubic box filled with polymers with
the same cross-linking density as the particle. Then, the pressure of the
cross-linked polymer box is equilibrated independently in $x$,$y$ and $z$ directions
in order to relax stresses to zero in all directions. 
Then the box is stretched in one direction by imposing a very small amount of
strain, typically $e_s \leq 0.02$. 
The length of the box is kept fixed in the
stretched condition and the stress $\sigma_s$ in that direction is measured.
The lateral stresses are kept equal to zero during the stretching and
relaxation stages, so that one directly probes the Young's modulus as

\begin{equation} \label{s-e}
 E_{\Delta t}=\frac{{\sigma_{s}}_{\Delta t}}{e_s},
\end{equation}
where $\Delta t$ shows the time interval over which stress is averaged. Fig.~\ref{f:E} presents the normalized value of the Young's modulus for different intervals of averaging. In molecular dynamics, stress exhibits much larger fluctuations as compared to velocity and density, since it is directly obtained from integrating rapidly changing interatomic forces (see e.g. Razavi et al. \citep{Razavi2014a}). In our simulations, the final value of elasticity is obtained by averaging the values of elasticities for $50,000\tau$ to $100,000\tau$ measurement periods, where we note that the averaging needs to be longer as the stiffness of the gel decreases.

\subsubsection{Surface tension measurement.~~}\label{int_ten}
The surface tension of the particle-liquid interface is measured using the Kirkwood-Buff formula for a planar interface \citep{Kirkwood1949}.

\begin{figure}[h]
\centering
  \includegraphics[height=7.5cm]{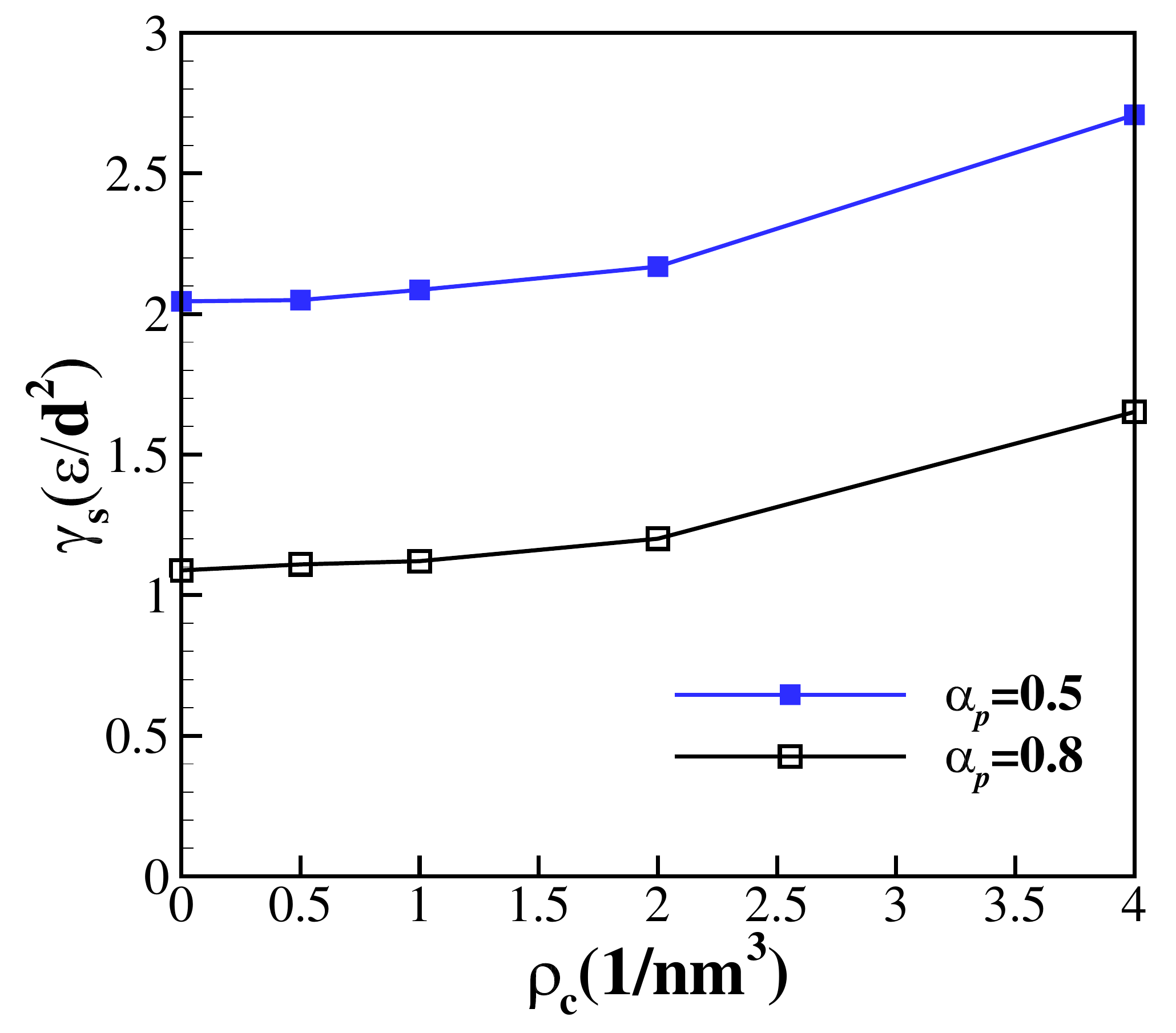}
\caption{Variation of the surface tension of the gel-liquid interface, $\gamma_s$, with the density of the cross-linking for two sets of interaction parameters: blue symbols are for $\alpha=0.5$, and  $\alpha_p=0.4$ and black symbols are for $\alpha=0.8$, and  $\alpha_p=0.8$. The planar interface is prepared inside a $120d \times 120d \times 120d$ box. After $10000\tau$ of equilibration, the surface tension is measured during a $20000\tau$ measurement period.} 
\label{f:gamma}
\end{figure}

This is done by creating a seperate setup with a planar interface between the
liquid phase and the cross-linked polymer chains. Fig.~\ref{f:gamma} shows how
the resulting surface tensions vary with cross-linking density. At small
cross-linking densities ($\rho_c \leq 1$), the surface tension increases weakly
with the cross-linking density and its value is very close to the liquid
droplet case which has no cross-linkings. As $\rho_c$ increases, its effect on
the surface tension becomes more significant.

\section{Results}\label{s:results}

\subsection{Partially wetting particles}\label{s:lens}

\begin{figure}[h]
\centering
  \includegraphics[height=7.5cm]{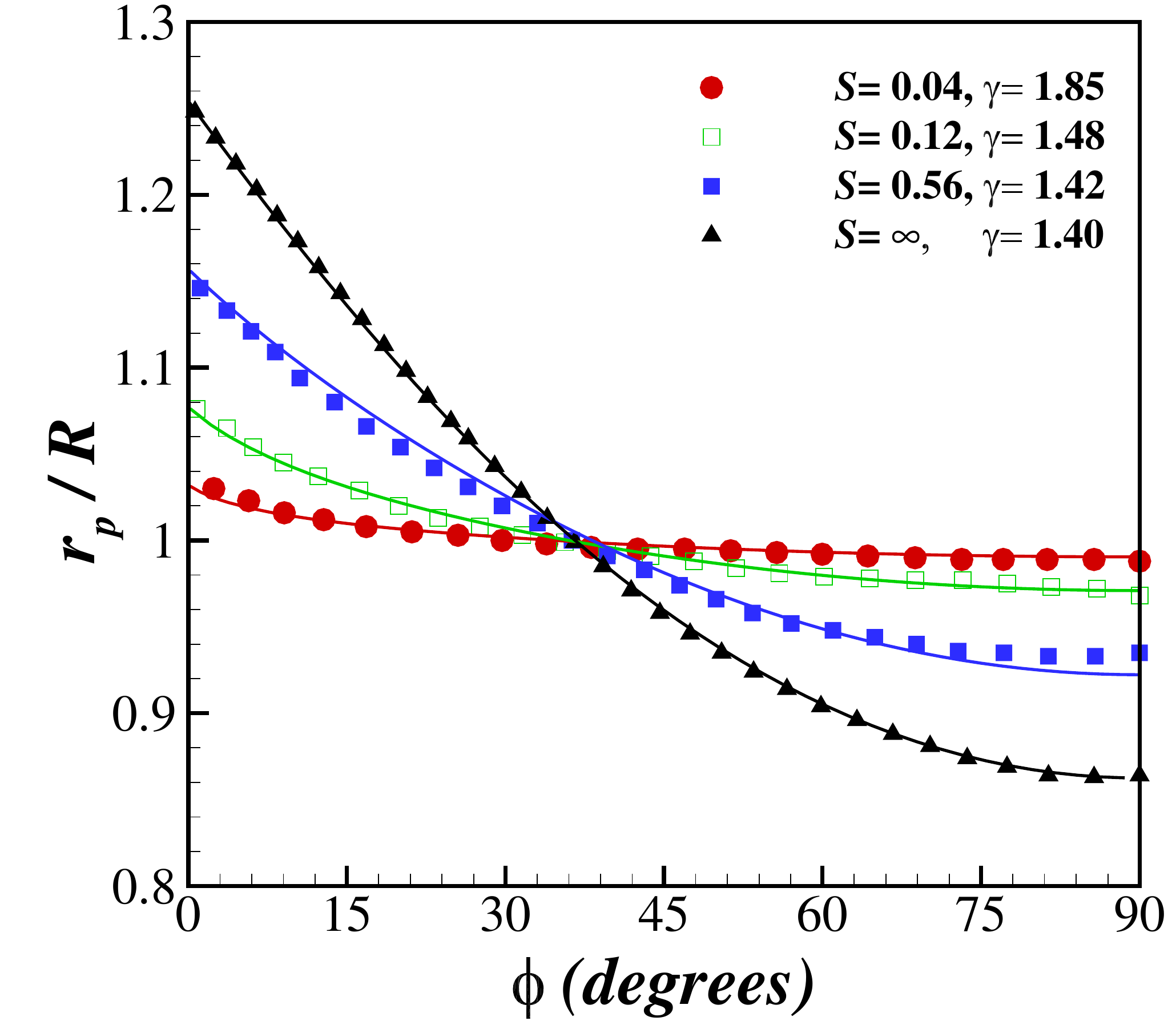}
  \caption{Shape of the particle in polar coordinates $(r,\phi)$. Symbols show the results of the molecular dynamics simulations. Solid lines are the theoretical predictions obtained from Eq. (\ref{eq:disk}). For $S=\infty$, a liquid droplet at a fluid interface, the line corresponds to a spherical cap shape.}
\label{f:particle_shape}
\end{figure}

We first consider simulations in the partially wetting regime, for which $\gamma > 1/2$. In addition, we focus on situations where the contact angle is fairly large ($\theta_s \geq 70^\circ$), so that deformations are small and fall within the linear regime. The key result is shown in Fig.~\ref{f:particle_shape}, where we characterise the particle shape by the radius $r_p$ as a function of the polar angle $\phi$, for a softness varying from $S=0.04$ (fairly rigid, high cross-linking density) to $S=\infty$ (liquid). Theoretical curves are shown as the solid lines and the results of the molecular dynamics simulations are shown using symbols. Except for the droplet ($S=\infty$) which has a spherical cap shape, theoretical curves are computed directly from  Eq. (\ref{eq:disk}). The agreement between the macroscopic theory and simulation is excellent, and emphasises the strength of the continuum approach even for discrete and fluctuating systems at small scales.

A few observations are of interest. For $S=0.04$, the deformation with respect to an undeformed particle is restricted to a small region close to the contact line, up to about $ \phi \sim 5^o$. The range of the deformation increases with $S$, and extends to the entire droplet at large $S$. Also the magnitude of the deformation increases, and as a consequence of the large deformations one observes some small deviations from the linear theory (e.g. for the blue squares). In the limit of very large softness, however, the elasticity drops out of the problem altogether and one finds again a perfect agreement with the purely capillary shape of a liquid droplet. 

\begin{figure}[h]
\centering
\includegraphics[height=7.5cm]{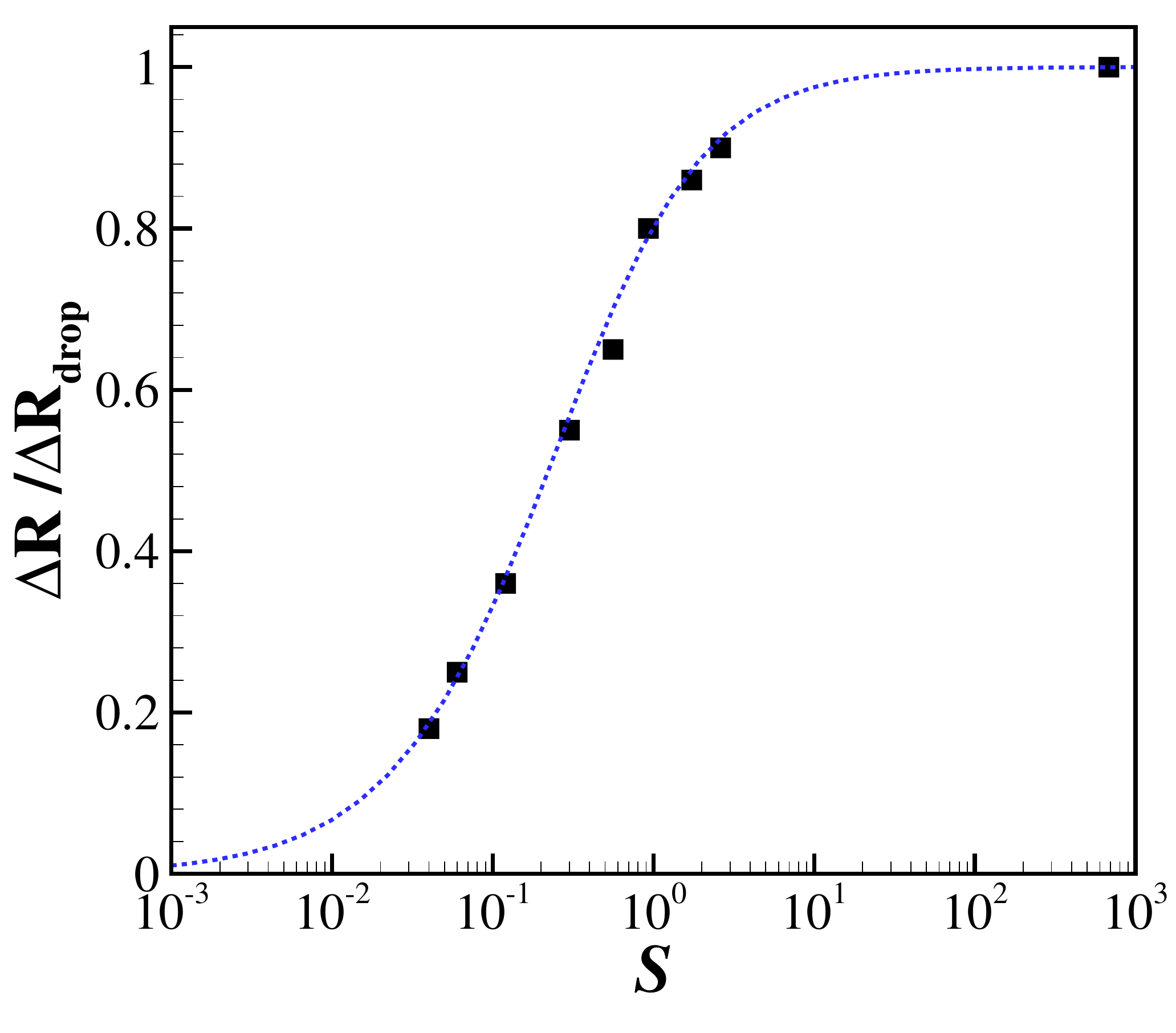}
\caption{Maximum radius of an elastic particle $\Delta R$ (normalised by that of a liquid drop $\Delta R_{drop}$), as function of softness $S$. Symbols correspond to molecular dynamics simulations, the solid line corresponds to the prediction given by (\ref{eq:DRS2}). The far right data point corresponds to a liquid droplet $(S=\infty)$. The radial extension increases continuously with $S$, and saturates at the value of a liquid droplet.  
}
\label{f:DR}
\end{figure}

Next we investigate the maximum deflection $\Delta R$ in order to further quantify the particle deformation. An important prediction of the macroscopic theory is that the scaled deflection $\Delta R/ \Delta R_{drop}$ is only a function of the softness $S$, and not of $\gamma$. The prediction (\ref{eq:DRS2}) is compared directly to the molecular dynamics simulation in Fig.~\ref{f:DR}, as solid line and symbols respectively. Again, a perfect agreement is found between the continuum theory and molecular simulations. At low $S$ the deformation radius increases $\sim S \log (S)$, as shown by (\ref{eq:log}), while at large $S$ the deflection saturates to the value known for a purely liquid drop $\Delta R_{drop}$. As expected, the crossover between the regimes of small and large deformations arises when $S \sim 1$. 
 
\begin{figure}[h]
\centering
  \includegraphics[height=7.5cm]{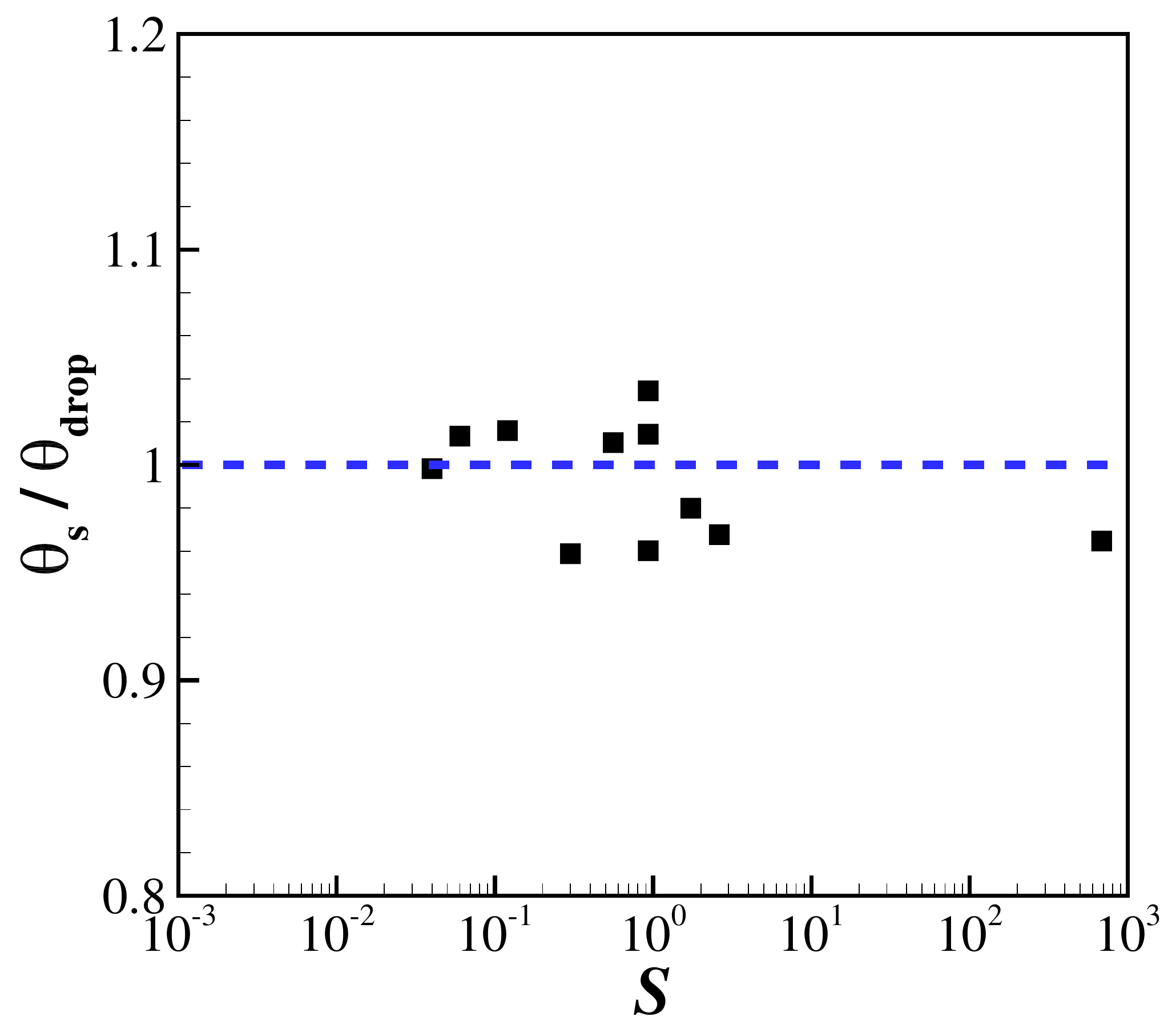}
  \caption{Variation of the normalized contact angle $\theta_s/\theta_{drop}$
with the softness. The solid line shows the theoretical value
($\theta_s=\theta_{drop}$) and symbols are the results from the molecular
dynamics simulations for $\theta_s$ between
$40^\circ$ and $82^\circ$. The simulation results fall within the $4\%$ of the theoretical values. The scatter in the data is because the contact angle does not correlate with the softness.}
\label{f:Qs}
\end{figure}

Finally, we can verify the prediction of the macroscopic analysis in section
\ref{s:theory} that the contact angle is determined by the surface tensions
according to the Neumann balance, regardless of the Young's modulus of the
particle.  In Fig.~\ref{f:Qs}, we plot the contact angle measured in our
simulations $\theta_s$ rescaled by the Neumann angle $\theta_{\rm drop}$. The
ratio is indeed very close to unity, over the entire range of softness -- even
for the stiffest particles -- meaning that the contact angle of the solid is
governed by the surface tensions. 
Data in Fig.~\ref{f:Qs}
covers contact angles from 40 to 82 degrees, meaning that the Neumann
balance is valid even for the contact angles much smaller than $90^\circ$.

\subsection{Completely wetting particles}\label{s:wetting}

\begin{figure}[h]
\centering
  \includegraphics[height=9cm]{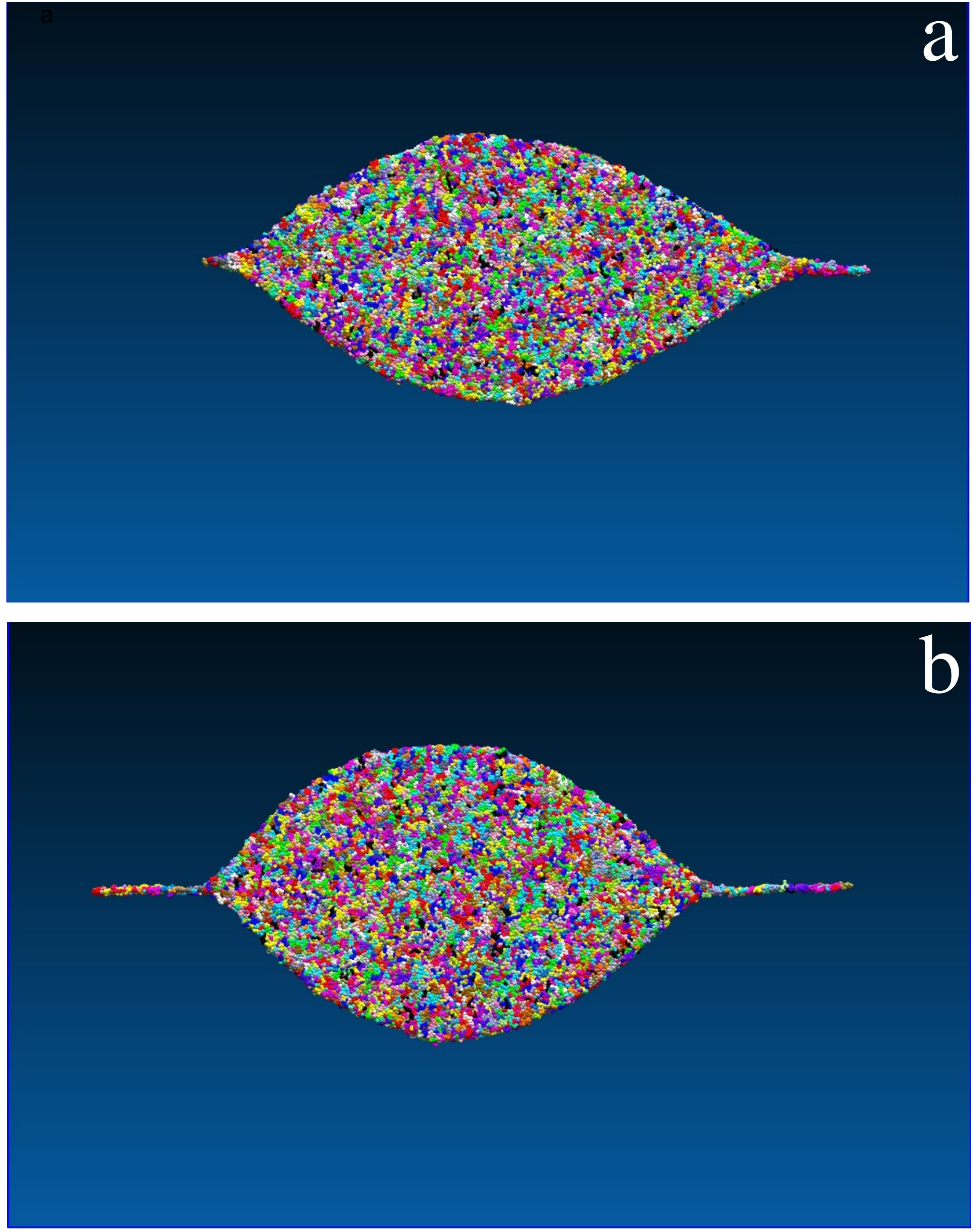}
  \caption{Effect of the cross-linking topology on the deformation of a
particle in the complete wetting regime. (a) Snapshot of a particle without any
free chains. The particle stretches until it reaches a final size. (b) Snapshot
of a particle with $30\%$ free chains. Free chains are pulled out of the
particle producing a thin liquid film. Both particles have identical
thermodynamic properties, $S=0.45$ and $\gamma=0.37$.}
\label{f:topology}
\end{figure}

For the partially wetting regime we have seen that the pulling force of the fluid interface is resisted by both the surface tension of the particle and its bulk elasticity: the capillary balance determines the geometry near the contact line, while at distances larger than $\gamma_s/E$ the bulk elasticity is predominant. An interesting paradox arises when the surface tension of the solid is not sufficiently strong to balance that of the fluid-fluid interface. For the uncross-linked polymers this leads to a wetting transition, leaving a thin film at the interface, but this is clearly prohibited when the polymer chains are strongly cross-linked. As we will argue in section \ref{s:discussion}, the situation of the complete wetting appears to be the generic case for microgel particles.
In Fig.~\ref{f:topology}, we present two cases with equal Young's modulus and ratio of surface tensions, i.e. identical $S$ and $\gamma$, but the bath surface tension is so high that both particles are in the complete wetting regime ($\gamma < 1/2$). Both particles also have an identical homogeneous density of cross-links, but the difference between panel (a) and (b) is the topology of the cross-linking network: in the upper panel \emph{all} polymeric chains are strongly entangled, while lower panel there are free chains in the particle that are not cross-linked. Note that even the spatial distribution of the links is similar between the two cases. 
The snapshot in Fig.~\ref{f:topology}b shows that free chains get pulled out of the particle and produce a polymer film at the interface. The particle without the free chains, by contrast, is very strongly stretched (Fig.~\ref{f:topology}a) -- yet due to the finite extensibility of the FENE-polymers, the particle remains at a finite size. In this case, the excess force near the contact line $\gamma_b - 2\gamma_s$ is balanced by the strong nonlinear elasticity of the cross-linked gel.
Furthermore, there is an asymmetry between the left and right hand side of the particle in Fig.~\ref{f:topology}(a) which also could be observed for the soft particle in complete wetting regime in Fig.~\ref{f:collage}. This shows that for particles in the complete wetting regime, minor asymmetry in the cross-linking can affect the symmetry of the particle, highlighting the role of microscopic properties on the shape of the particle, while for particles in the partially wetting regime, surface tension undermines such effects. This result illustrates that (i) the completely wetting regime cannot be captured by linear elasticity, and (ii) contrarily to the partially wetting case, molecular details become of key importance to describe the equilibrium shape.

Finally, let us investigate the influence of the softness on the particle shape in the case where no free chains are present. This is revealed in Fig. 11, where we show the relative extension of the particle $\Delta R/R$ as a function of $S$. The snapshots at small $S$ reveal how a thin region near the contact line is pulled out of the particle: due to the complete wetting condition, the pulling force of the bath surface tension cannot be balanced by the surface tension of the gel, and hence no Neumann triangle can be formed. As a consequence, the particle elasticity must provide a highly localised stress at the contact line, which is achieved here be forming a cusp-like tip. Clearly, this feature is beyond linear elasticity and the linear theory does not suffice to capture the complete wetting regime. Our simulations show that the size of this tip-region increases as the particle gets softer, and in the range where $S \sim 1$ it invades the entire particle. In all cases, the strongly nonlinear deformation provides an elastic force that can compete with the surface tension force near the contact line. This is to be contrasted with the partially wetting case where elasticity does not play a role at the contact line.

\begin{figure*}[h!t]
\centering
  \includegraphics[height=10cm]{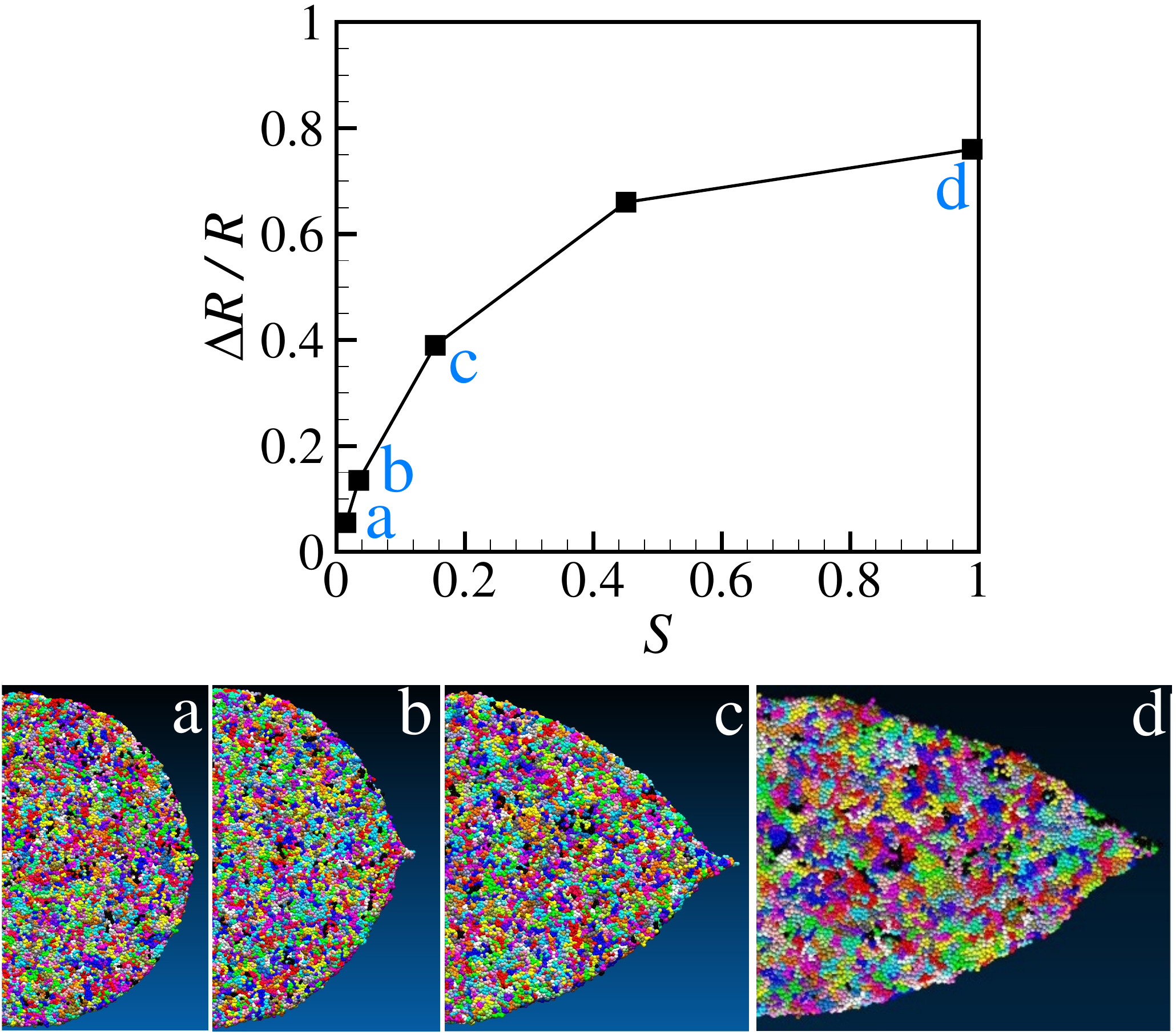}
  \caption{
  Maximum radius $\Delta R$ (normalised by the initial drop radius $R$), as a function of $S$ in the case of complete wetting. All polymers are cross-linked so that the particle reaches an equilibrium with finite deformation. All particles are subjected to the same pulling force and are in the complete wetting regime. For numerical reasons, the largest value of $S$ was achieved by a slightly smaller particle.}
\label{f:S}
\end{figure*}

\section{Discussion}\label{s:discussion}

We have investigated the shapes of deformable, elastic particles when adsorbed at a fluid-fluid interface. Apart from the elasticity of the particle, we have demonstrated that one crucially needs to consider the surface tension of the particle. In particular, one needs to distinguish cases of partial wetting and complete wetting. It was found that partially wetting particles simulated by Molecular Dynamics are accurately described by the continuum elasticity theory, while for the case of complete wetting, molecular details such the cross-linking topology turn out to be important. While the present paper has focused on symmetric particles, with equal affinity to the two liquid phases, it is interesting to consider more general values for the surface tensions. The case of partial wetting is reached when the sum of surface tensions of the solid $\gamma_{s1}$ and $\gamma_{s2}$, respectively with fluid ``1" and ``2", are larger than that of the bath $\gamma_b$. At the same time, for adsorption to be energetically favourable with respect to the particle being in phase ``2", one requires $\gamma_{s1}$ to be lower than $\gamma_{b}+ \gamma_{s2}$. Partially wetting and adsorption is thus reached when the following inequality is satisfied:

\begin{equation}\label{eq:inequality}
\gamma_b - \gamma_{s2} < \gamma_{s1} < \gamma_{b} + \gamma_{s2}.
\end{equation}

It is interesting to consider this condition in the context of microgel particles. These are hydrogels that have a nonzero but small surface tension
with respect to the water phase in which they are immersed. Denoting the water as phase ``2", we typically find that the gel-water surface tension $\gamma_{s2}$ is much smaller than the other surface tensions. In terms of the inequality (\ref{eq:inequality}) this implies that the window for the partial wetting regime of adsorbed particles is extremely small, and that one generically expects the particle to be in completely wetting situations similar to what can be seen in our Fig.~\ref{f:topology}. 
Experimentally, it has remained challenging to explain the core-corona like shape found for the microgels \cite{Destribats2014,Deshmukh2014a,Monteillet2014}. 
As is shown in Fig.~\ref{f:topology}, the core-corona shape could happen either due to the deformation of the entangled polymeric network of the microgel particles or due to the adsorption of the individual polymeric chains. 
For a typical microgel particle, with a Young's modulus of 100 KPa \citep{Burmistrova2011}, a particle-oil surface tension of 50 mN/m, and a typical radius of a 0.5 micron, the softness parameter will be of order unity. Given that experiments are likely to be in the complete wetting regime, according to Fig.\ref{f:S}, the expected deformation is therefore of the order of the particle radius consistent with the experiments \citep{Geisel2012,Deshmukh2014a}. The important observation from comparing the snapshots of Fig.\ref{f:S} is that for a particle without free chains, the size of the corona (cuspy shape close to the contact line) which is clearly present for small $S$ (snapshots $a$ to $c$) vanishes for the particle whose softness is close to one (snapshot $d$) which is the typical experimental value. This shows that for $S\approx 1$, the deformation spans over the entire particle shape instead of producing an elongated cusp close to the contact line. If we compare this observation to the experimental observations that the corona size is comparable to the particle core size \citep{Geisel2012,Deshmukh2014a}, we can conclude that the details of adsorption of individual polymer chains play an important role in producing a core-corona shape for microgels at a fluid interface. It should be noted that in this study as the first step to model the behavior of microgel particles at a fluid interface, particles are considered to be in the dry state which means that the solvent particles do not penetrate into the microgels and the effect of particle swelling is ignored. It would be interesting to extend the present work to incorporate the effect of particle swelling which is usually present in experiments.

\section{Conclusions}
The shape of a soft particle at a fluid interface is studied both numerically and analytically. It is shown that the surface tension of the particle, expressed in terms of the dimensionless parameter $\gamma$ plays an important role in characterizing the particle shape.
For $\gamma \geq 1.4$, i.e. particles whose surface tension is large compared to the fluid surface tension, it is analytically shown that the contact angle of the particle is equal to the Neumann contact angle of a liquid droplet with similar $\gamma$. In addition, an analytical formula is derived for the particle shape and it is shown that it scales with the shape of the liquid droplet with similar $\gamma$ where the scaling coefficient is only a function of the particle's softness $S=\frac{\gamma_s}{ER}$.
The derived analytical equation for the particle shape and contact angle is examined using the coarse grained molecular dynamics simulations of particles made out of entangled polymeric chains and it is shown that the analytical and numerical results are in a perfect agreement.
For $ 0.5 \leq \gamma < 1.4$, molecular dynamics simulations show that the analytical formula for the particle shape loses its accuracy as it is expected but the contact angle of the particle is still equal to the Neumann angle.
For $ \gamma < 0.5$, which considers the particles in the complete wetting regime, it is also shown that the shape of the particle depends on the microscopic details of the cross-linking. For example, two particles with similar macroscopic properties but different number of free chains produce different final shapes. 
Using the results of the particle in the complete wetting regime, it is argued that the core-corona structure which is observed in the majority of the microgel experiments appears due to the adsorption of free chains inside the particle to the interface.

\section{Acknowledgements}
We are grateful to Martien Cohen-Stuart, Jasper van der Gucht and Joris Sprakel for many discussions on the adsorption of microgel particles and Joost Weijs for making the GROMACS setup. We acknowledge financial support from NWO through VIDI Grants No. 11304 (JS) and 10787 (JH), and financial support from ERC (the European Research Council) through Consolidator Grant No. 616918 (JS).

\appendix

\section{Appendix} 
As we are dealing with a two-dimensional problem, we can exploit the Airy stress function $\Phi$ and solve the biharmonic equation $\nabla^4 \Phi=0$ \cite{Musk}. The corresponding stress and displacement fields can be determined as partial derivatives of $\Phi$. In polar coordinates, we can proceed by a Fourier expansion using the classical solution by Michell. The up-down symmetry is such that we only need to consider Fourier modes of the type $\cos n\phi$, where $n=0,2,4, \cdots$. The isotropic solution $n=0$ comes with powers $r^2, r^2 \ln r, \ln r$. The $n=2,4,6, \cdots$ solutions to the biharmonic equation come with radial powers of the type $r^n, r^{n+2}, r^{-n}, r^{-n+2}$. Demanding regularity of stress and strain at $r=0$, we cannot allow for $r^{-n},r^{-n+2}$ or the logarithmic terms, so that:
\begin{equation}\label{eq:airy}
\Phi(r,\phi) = A r^2 +  \sum_{n=2}^\infty \cos n\phi \left(a_n r^n + b_n r^{n+2}\right),
\end{equation}
The corresponding displacements (taken at the free surface $r=1$) become \cite{Musk}:

\begin{eqnarray}\label{eq:ur}
u_r(1,\phi) &=& \frac{3 S}{2\gamma} \sum_{n=2}^\infty \cos n\phi \left(-n a_n  -n b_n \right), \\
u_\phi(1,\phi) &=& \frac{3 S}{2\gamma} \sum_{n=2}^\infty \sin n\phi \left(n a_n  +(n+2) b_n \right),
\end{eqnarray}
where $S$ is the anticipated dimensionless parameter to characterise the softness and $\gamma$ is the ratio of the surface tensions. Due to the incompressibility ($\nu=1/2)$ there is no isotropic contribution to the displacement. Likewise, one expresses the stress at $r=1$ from the expansion (\ref{eq:airy}) as

\begin{eqnarray}\label{eq:stress1}
\sigma_{rr}(1,\phi) &=& 2A + \sum_{n=2}^\infty \cos n\phi \left[ (n-n^2)a_n \right. \nonumber \\
&& \left. + (n+2-n^2)b_n \right], \\ 
\sigma_{r\phi}(1,\phi) &=& \sum_{n=2}^\infty n\sin n\phi \left[ (n-1)a_n + (n+1)b_n \right].\label{eq:stress2}
\end{eqnarray}

The final step is to expand the boundary conditions (\ref{eq:bc1},\ref{eq:bc2}) in a Fourier series and compare this to the representation (\ref{eq:stress1},\ref{eq:stress2}). This will give two equations for the coefficients $a_n,b_n$. In particular, the normal stress boundary condition becomes, using also (\ref{eq:ur})

\begin{equation}
\sigma_r(1,\phi) = \frac{c_0}{2} + \sum_{n=1}^\infty c_n \cos n \phi,
\end{equation}
with coefficients 

\begin{equation}
c_0 = \frac{2}{\pi}, \: c_n = \frac{2}{\pi} + \frac{3}{2} S \left[ n(n^2-1) (a_n+b_n) \right] \: n=2,4,6, \cdots, 
\end{equation}
while $c_n=0$ for odd $n$. Solving the coefficients $a_n,b_n$ finally gives

\begin{equation}
a_n = \frac{1}{\pi(1-n)\left[ 1+ \frac{3}{2} S n \right]}, \quad b_n = \frac{1}{\pi(1+n)\left[ 1+ \frac{3}{2} S n \right]}.
\end{equation} 

The displacement field can be summarized in explicit form:

\begin{eqnarray}
u_r(r,\phi) &=&  - \frac{3 S}{2\gamma\pi} \sum_{k=1}^\infty  \cos 2k\phi \left[ r^{2k-1} \left(\frac{2k}{(1-2k)\left( 1+ 3 S k  \right)}\right) \right. \nonumber \\
&& + \left. r^{2k+1} + \left(\frac{2k}{(1+2k)\left( 1+ 3 S k \right)}\right)   \right]. \\
u_\phi (r,\phi) &=& \frac{3S}{2\gamma\pi} \sum_{k=1}^\infty  \sin 2k\phi \left[ r^{2k-1} \left(\frac{2k}{(1-2k)\left( 1+ 3 S k \right)}\right) \right. \nonumber \\
&& + \left. r^{2k+1} \left(\frac{2k+2}{(1+2k)\left( 1+ 3 S k \right)}\right)   \right].
\end{eqnarray}

The shape of the deformed particle is thus given by

\begin{eqnarray}\label{eq:diskshape}
r_{\rm p}(\phi) &=& 1+ u_r(1,\phi) \nonumber \\
&=&  1 - \frac{3S}{2\gamma\pi} \sum_{k=1}^\infty \frac{4k \cos 2k\phi}{(1-4k^2)(1+3 S k)}.
\end{eqnarray}

\section*{References}

\providecommand*{\mcitethebibliography}{\thebibliography}
\csname @ifundefined\endcsname{endmcitethebibliography}
{\let\endmcitethebibliography\endthebibliography}{}


\end{document}